\documentclass[rsi,
 aip,
 amsmath,amssymb,
 reprint,%
]{revtex4-1}

\usepackage{graphicx}
\usepackage{dcolumn}
\usepackage{bm}

\usepackage{amsmath,amssymb,amsfonts}%
\usepackage{xcolor}%
\usepackage{booktabs}

\usepackage[utf8]{inputenc}
\usepackage[T1]{fontenc}
\usepackage{mathptmx}
\usepackage{etoolbox}

\makeatletter
\def\@email#1#2{%
 \endgroup
 \patchcmd{\titleblock@produce}
  {\frontmatter@RRAPformat}
  {\frontmatter@RRAPformat{\produce@RRAP{*#1\href{mailto:#2}{#2}}}\frontmatter@RRAPformat}
  {}{}
}%
\makeatother
\begin{document}

\preprint{AIP/123-QED}

\title[Electron selector]{High-charge collimated and energy-selected laser-driven MeV electron beams produced by magnetic selection}
\author{I. Cohen}
 \affiliation{Department of Physics, Technion, Haifa, 32000, Israel}

\author{I. Slabu}
\affiliation{``Horia Hulubei'' National Institute for Physics and Nuclear Engineering, RO-077125 Bucharest-Magurele, Romania}
\affiliation{University of Bucharest, Romania}

\author{Q. Peysson}%

\affiliation{ 
LULI - CNRS, CEA, UPMC Univ Paris 06 : Sorbonne Universit\'e, Ecole Polytechnique, Institut Polytechnique de Paris - F-91128 Palaiseau cedex, France
}%
\author{S. Dorard}%

\affiliation{ 
LULI - CNRS, CEA, UPMC Univ Paris 06 : Sorbonne Universit\'e, Ecole Polytechnique, Institut Polytechnique de Paris - F-91128 Palaiseau cedex, France
}%
\author{Y. Abe}
\affiliation{Graduate School of Engineering, The University of Osaka, 2-1
Yamadaoka, Suita, Osaka, 565-0871, Japan}
\author{J. Béard}
\affiliation{Laboratoire National des Champs Magnétiques Intenses, LNCMI-CNRS, EMFL, Université Grenoble-Alpes, Université Toulouse 3, INSA Toulouse, F-31400 Toulouse, France}
\author{T. Moraine}
\affiliation{Laboratoire National des Champs Magnétiques Intenses, LNCMI-CNRS, EMFL, Université Grenoble-Alpes, Université Toulouse 3, INSA Toulouse, F-31400 Toulouse, France}
\author{S.N. Chen}
\affiliation{Light Stream Labs LLC, USA, Palo Alto, CA 94306}
\author{A. Chessa}
\affiliation{LULI - CNRS, CEA, UPMC Univ Paris 06 : Sorbonne Universit\'e, Ecole Polytechnique, Institut Polytechnique de Paris - F-91128 Palaiseau cedex, France}
\author{K. Iida}
\affiliation{Graduate School of Engineering, The University of Osaka, 2-1
Yamadaoka, Suita, Osaka, 565-0871, Japan}
\author{P. Kempski}
\affiliation{Instituto Superior Técnico, Universidade de Lisboa, Av Rovisco Pais, 1049-001 Lisboa, Portugal}
\author{Y. Kuramitsu}
\affiliation{Graduate School of Engineering, The University of Osaka, 2-1
Yamadaoka, Suita, Osaka, 565-0871, Japan}
\author{H. Kusano}
\affiliation{Graduate School of Engineering, The University of Osaka, 2-1
Yamadaoka, Suita, Osaka, 565-0871, Japan}
\author{F. Nikaido}
\affiliation{Graduate School of Engineering, The University of Osaka, 2-1
Yamadaoka, Suita, Osaka, 565-0871, Japan}
\author{M. Ruszkowski}
\affiliation{Department of Astronomy, University of Michigan, 1085 S. University Ave., 323 West Hall, Ann Arbor, MI, 48109-1107, USA}

\author{K. Sakai}
\affiliation{National Institute for Fusion Science, 322-6 Oroshicho, Toki, Gifu 509-5292, Japan}

\author{N. Tamaki}
\affiliation{Graduate School of Engineering, The University of Osaka, 2-1
Yamadaoka, Suita, Osaka, 565-0871, Japan}
\author{O. Tesileanu}
\affiliation{``Horia Hulubei'' National Institute for Physics and Nuclear Engineering, RO-077125 Bucharest-Magurele, Romania}

\author{J. Fuchs}
\affiliation{%
Department of Physics, Technion, Haifa, 32000, Israel
}%
\affiliation{ 
LULI - CNRS, CEA, UPMC Univ Paris 06 : Sorbonne Universit\'e, Ecole Polytechnique, Institut Polytechnique de Paris - F-91128 Palaiseau cedex, France
}%
 \email{Julien.fuchs@technion.ac.il}

\date{\today}

\begin{abstract} 
We have developed a compact passive energy-selector for MeV-range electrons produced by irradiating solid targets by ultra-intense short-pulse lasers. The device allows for generating electron beams with a variable energy spread over a broad range of energies, from tens of keV to tens of MeV. 
Here we have demonstrated its use by producing electrons 
from solid targets in the MeV range and with a $\sim$10\% bandwidth
, thereby compensating the intrinsic broadband nature of the electrons produced from such source. Coupled with a pulsed magnetic field to further compensate the intrinsic large divergence of this source, it allows to produce a highly-collimated beam of narrow-band and ultra-fast electrons, suitable for a wide range of applications, e.g. radiation therapy or time-resolved electron probing. 
\end{abstract}

\maketitle

\section{\label{sec:level1}Introduction}

Fast (MeV) electron beams are used, or considered to be used, in a wide variety of applications, e.g. radiation therapy \cite{Renard2024}, electron radiography \cite{Merrill2007,Xiao2018}, electron diffraction \cite{Filippetto2022} or electromagnetic field radiography \cite{Schumaker2013}. For all, the use of lasers, and in particular short-pulse lasers \cite{Strickland1985}, have brought a new capability, namely ultra-high temporal resolution. The way these lasers have been  used 
was to either drive  RF photocathode of conventional accelerators, or to directly drive the acceleration of electrons in low-density plasmas. This can be done either through exciting wakefield plasma structures  \cite{Krushelnick2010}, or through direct laser acceleration \cite{Hazra2019,Cohen2024}. 
Note that, in general, compared to electron beams accelerated by conventional accelerators, laser-driven beams have much less repetition rate at present, limited by the laser driver, but they are conversely much more compact in size \cite{Guo2026}, due to the accelerating gradients in plasmas being at least four orders of magnitude higher than those in e.g. drift-tube linacs. A further advantage of laser-driven acceleration is the capability to produce high-charge beams, when conventional accelerators   face the issue that the electron spectrum per bunch are limited by space charge effects \cite{Zhu2015}. However, while wakefield acceleration produces mono-energetic electrons, these are  still accompanied by an additional spectral broadband component. In contrast, the  spectra of direct laser acceleration-produced electrons are broadband.


A way to circumvent the issue, and generate stable and controlled mono-energetic bunches, while keeping a short duration for the bunch, is to filter the spectrum in a controlled manner.
The present paper is focused on describing a method to generate both high-charge, controlled narrow-band and collimated electron beams in the MeV-range, i.e. to alleviate the limitations that were faced by all previous methods. It relies on coupling  passive and pulsed \cite{Bolaos2019} magnetic devices to electrons produced from solids. All this, combined, allows to energy-select in a stable manner and collimate MeV-range electrons. The intend is to bring an energy control capability to electrons laser-driven from solids, which are intrinsically broadband and divergent. The interest of using such a source of electrons, compared to driving the electrons from low-density plasmas, is that (i) they generate high-charge electrons, i.e. even after selection, the electron beams still have nC charge, and (ii) the energy selection and collimation, being induced by external devices, are unaffected by the shot-to-shot fluctuations of laser-plasma interactions. Note that although laser wakefield produced electrons can have also nC charge beams \cite{Chiu2004,Martelli2025}, they are mostly produced with pC charge \cite{Monzac2024}. 

Electrons accelerated by lasers from the surface of solid targets use  combined E and B fields of the laser which cannot penetrate into the target beyond a skin depth. These fields thus interact with the strong plasma gradients that are present at the target surface, leading to the  acceleration of bunches of electrons inward and outward from the solid \cite{Haines2009,Kluge2018}. 
Electron beams laser-accelerated from solids differ in characteristics from the electron beams produced from low-density gases: while the electrons accelerated from solids are characterized by large number of electrons per bunch (5~$\mu C$\cite{Rosmej2019}), since the process has a high efficiency (that can reach $\geq$~50$\%$\cite{Andreev2021,Cohen2024,Rosmej2019}, dependent on the laser intensity \cite{Haines2009}), they are also widely divergent (90° full divergence) \cite{Debayle2010,Green2008} and broadband (except for particularly tuned parameters \cite{Mordovanakis2009}). 

\begin{figure*}[hbtp]
\centering
\includegraphics[trim={0.0cm 1.3cm 0cm 0cm},clip, width=1\textwidth]{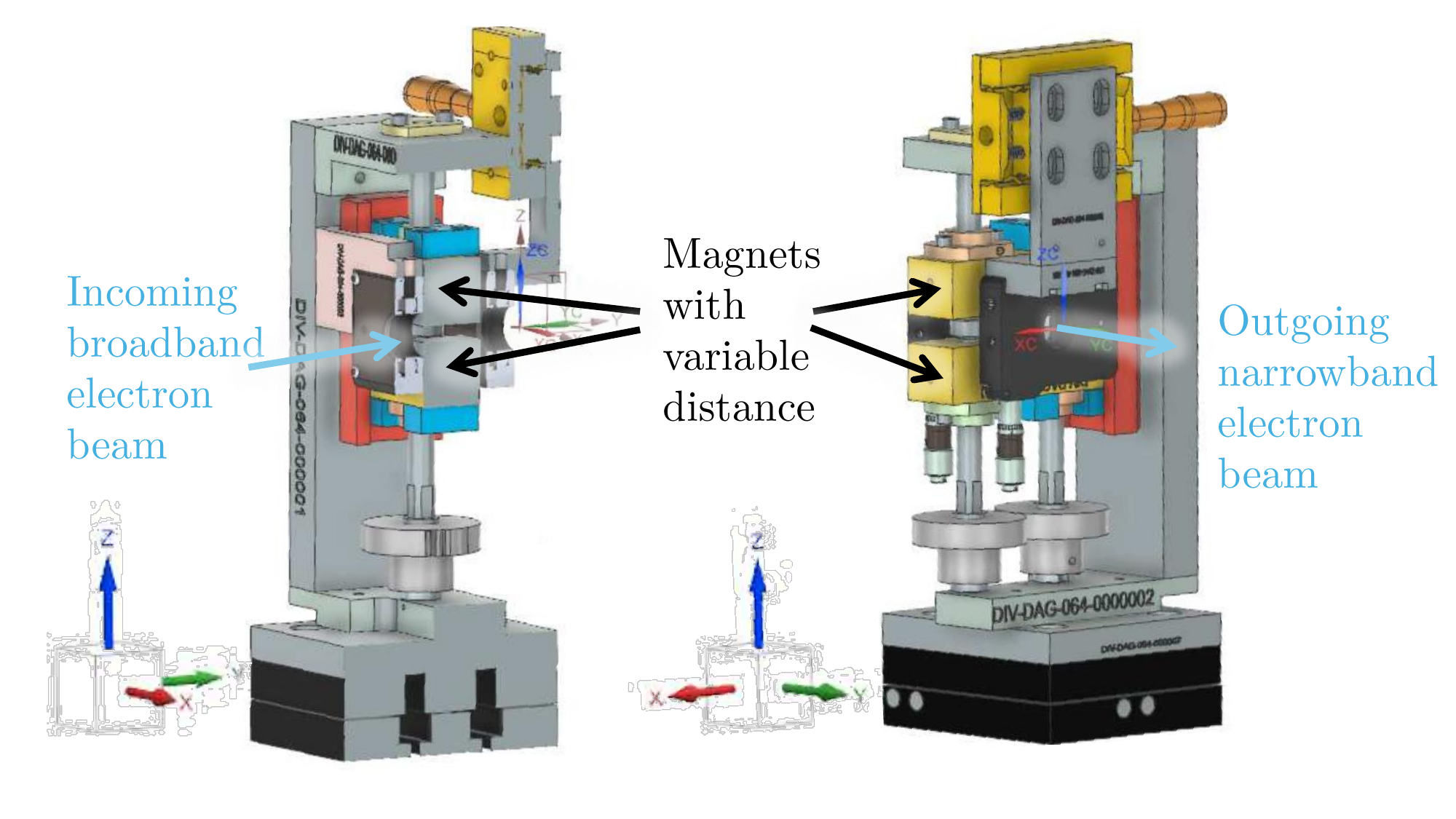}
\caption{\label{setup} Design and illustration of the passive electron energy selector. A broadband, wide-angle, source of electrons, usually driven by a high intensity laser, propagates  into a slit followed by 2 magnets which bend the electrons into another thick slit, allowing only electrons with a certain energy and propagation angle to pass through. The magnets are attached to a  stage that changes the distance between them, thus allowing to tune the magnetic field strength and thus the selected electron energies.}
 \label{fig:setup}%
 \end{figure*}

The 
paper is organized as follows: we describe the passive magnetic device, shown in Fig.\ref{fig:setup}, that induces the energy selection, its design and realization. We then present the experimental characterization of the energy-selection it allows for initially broadband electrons. We then show how it can be coupled to a pulsed magnetic field device in order to collimate the produced electron beams. Lastly, we show, as an example of application, how the whole system can be exploited to perform time-resolved measurement of electron transport in a plasma.

\section{Experimental setup and electron energy-selection}

\begin{figure}[hbtp]
\centering
\includegraphics[trim={0.0cm 0cm 0cm 0cm},clip, width=0.5\textwidth]{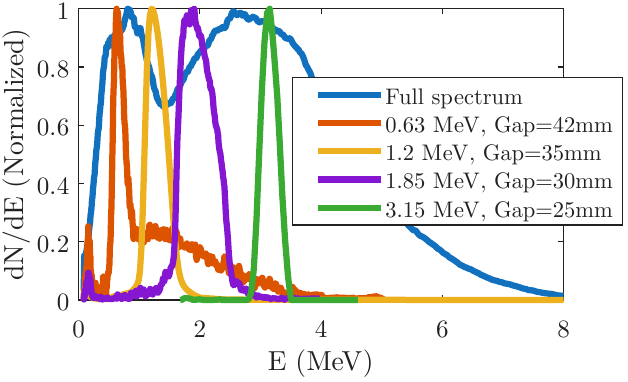}
\caption{ Electron spectra measured without the selector ("full spectrum") and with selector, for four different values of the selecting magnetic field, as given by the gap between the magnets and as stated in the legend. The electrons were measured using a magnetic spectrometer, described in Ref.~\cite{Cohen2026}, with an imaging plate \cite{Bonnet2013} placed on the detector plane.  }
 \label{fig:spectra}%
 \end{figure}

To produce the electrons, we use  the 1 ps duration, ~50 J energy short-pulse of LULI2000 \cite{Zou2008}. The laser is  focused to an elliptical spot having a 7.0 $\times$ 6.9 $\mu m^{2}$ full width at half maximum (FWHM), leading to  a ~$10^{19}$~W/cm$^2$ intensity. The beam is focused onto a 20 microns thick PET target. Following the interaction, a broadband electron beam is produced. It is  diagnosed by placing, downstream and in the laser axis (as the centroid of the electron beam follows the laser axis \cite{Rusby2015,Santala2000}), a spectrometer equipped with a permanent magnet of 2.5~T. As shown by the blue curve in Fig.\ref{fig:spectra}, the produced electrons are characterized by an exponential energy distribution at high energy, with a temperature of 0.85 MeV, and a plateau at low-energy in the  (0.1-3 MeV) range. 

\begin{figure}%
    \includegraphics[trim={0.0cm 0cm 0cm 0cm},clip, width=8cm]{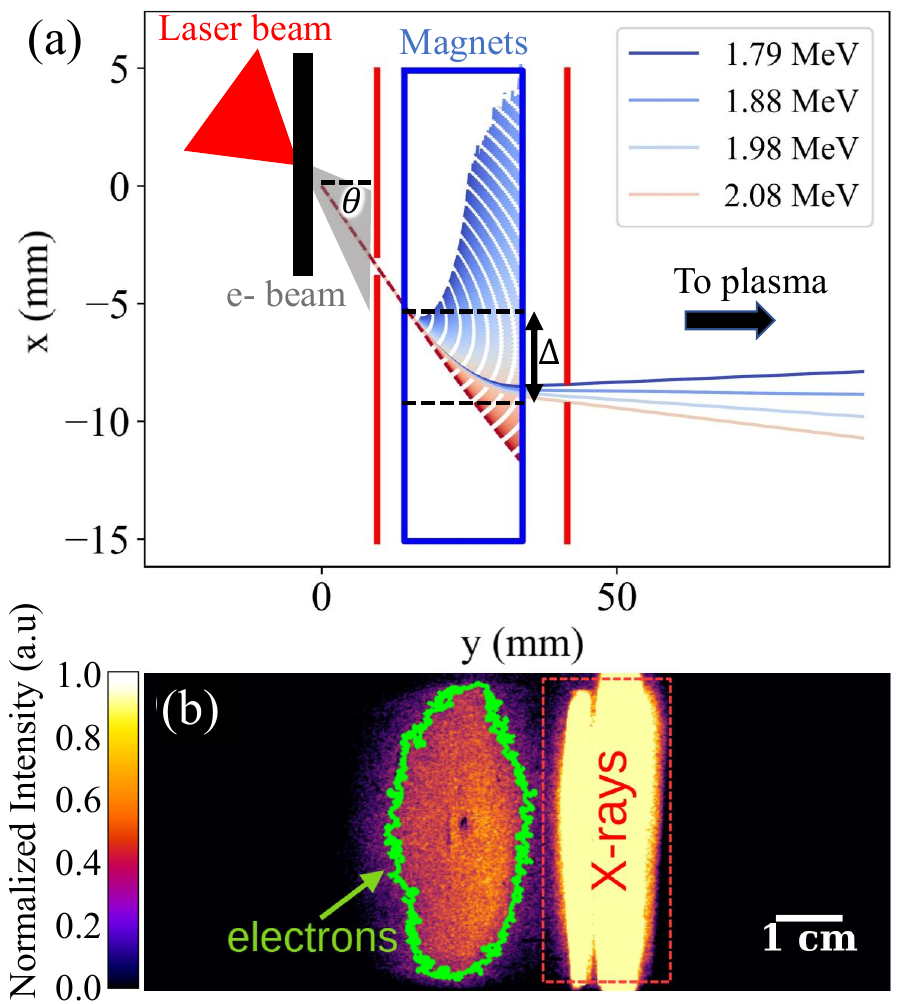} 
    \caption{(a) Analytically-calculated  electron trajectories within the selector for a magnetic field of 0.132 T at the center, corresponding to a gap between the magnets of 30 mm. The slits  are draw in red and the magnetic field zone in blue. The laser-target is located at x  = y = 0. All the electrons are emitted from the target with energies from 0.1 (blue dashed lines) to 15 MeV (red dashed lines). Only the discretized energies that pass the second slit are labeled.
    (b) Electron beam profile recorded 11 cm from the exit of the selector, along with the image (saturated line on the right of the spot) of the generated X-rays from the laser-target interaction, and which passes straight through the slits of the selector. The beam is collected on an imaging plate. The green line delineates the 1/e contour of the beam.}%
    \label{fig:RT_Simulation}%
\end{figure}

As illustrated in Fig.\ref{fig:RT_Simulation}a, the electrons are then sent into a passive  energy-selector. It relies on a simple design. An angularly narrow portion of the angularly broad \cite{Rusby2015,Santala2000} electron beam is selected, using a hard aperture, then sent into a dipole magnet to disperse the beam in energy. A second hard aperture is placed downstream, in order to energy-select the beam. The purpose of the first aperture is to carve out a beamlet within the broad cone of emission of electrons produced from the solid target. Within this beamlet, the electrons retain the full broadband spectrum they are characterised by at the source (see Fig.\ref{fig:spectra}). 
The relative positions of the two apertures determine the angle with which the electrons will be streaming in the free space downstream, as well as the average energy of the selected electrons. The width of these two apertures determines in turn the divergence of the exiting beam, as well as its energy bandwidth. An image of the electron beam exiting the selector, collected by an imaging plate positioned 11 cm downstream from the selector exit, is shown in Fig.\ref{fig:RT_Simulation}b. This beam is obviously divergent, but is now energy-selected, as will be detailed below.

\begin{figure}[hbtp]
\centering
\includegraphics[trim={0cm 0cm 0cm 0cm},clip, width=0.5\textwidth]{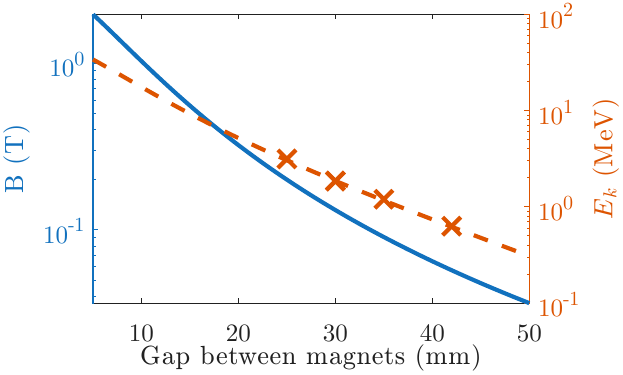}
\caption{Analytical calculation of the magnetic field strength at the center of the selector vs the gap between the magnets (blue line, left axis, as given by Eq.~\ref{B_analytical}) and of the selected electron energy vs the gap (red dashed line, right axis, as given by Eq.~\ref{Ek_analytical}) with measurements (crosses, corresponding to the peak energies of the four selected beams shown in Fig.\ref{fig:spectra}).} 
 \label{fig:calculations}%
 \end{figure}

The energy selection is entirely determined by the geometry of the arrangement and the characteristics of the magnets. We set the selector such that the incident angle of the electron beamlet injected in it should be at 110° from the exit axis of the electrons, as illustrated in Fig.\ref{fig:RT_Simulation}a. This being fixed, the selection in energy of the electrons will depend on the path they travel within the magnetic field and on  the strength of the latter. The magnetic field results from the addition of the individual fields induced by the individual magnets set in the selector (see Fig.\ref{fig:setup}). The magnetic field at the center of the selector, for a single magnet, can be  calculated analytically and is given in Eq.~\ref{B_analytical}. There, $B_r$ is the remanence magnetic field, $a$ is the length of the magnet, $b$ is the width of the magnet, $l$ is the height of the magnet, and $d$ is the distance from the magnet surface. 

\begin{equation}
\begin{split}
    B=\frac{B_r}{\pi} \left[ atan\left(\frac{ab}{2d\sqrt{a^2+b^2+4d^2}}\right) \right. \\
    \left. -atan\left(\frac{ab}{2(d+l)\sqrt{a^2+b^2+4(d+l)^2}}\right) \right]
    \label{B_analytical}
\end{split}
\end{equation}
The blue curve in Fig.\ref{fig:calculations} illustrates how the strength of the magnetic field at the center of the selector varies as a function of the gap between the magnets.
The selection of electron energies at the selector exit is given by Eq.~\ref{Ek_analytical}. This equation relates the geometry of the selector to the magnetic field at its center and the  electron energy. 
\begin{equation}
    E_k=m_ec^2\left(\frac{1}{\sqrt{1-(\frac{eB}{m_ec}\frac{\Delta}{1-\cos{\theta}})^2}}-1\right)
    \label{Ek_analytical}
\end{equation}
 In Eq.~\ref{Ek_analytical}, $\Delta$ is the relative position shift of the electrons in the magnetic field along the x-axis (see Fig.\ref{fig:RT_Simulation}a), and $\theta$ is the angle between the entrance and exit direction of the electrons, as shown in Fig.~\ref{fig:RT_Simulation}a. In our design $\theta=20^\circ$, and $\Delta=3.5mm$.
 
 Fig.\ref{fig:spectra} demonstrates the effective energy-selection that is performed onto the initially broadband electron beam, for different values of the magnetic field at the center (varied by adjusting the spatial gap between the magnets, see  Fig.\ref{fig:setup}). Note that here the electron spectra are measured with the same spectrometer as used to characterized the unselected beam, but with the spectrometer being rotated to be along the exit axis of the selector (the y-axis). The peak energies of the selected electrons, as recorded in Fig.\ref{fig:spectra}, are summarized in Fig.\ref{fig:calculations}, which shows that the experimental values of the selected electron energies are in excellent agreement with the analytical model given by Eq.~\ref{Ek_analytical}.
 
  \section{Collimated beam characteristics}
 
\begin{figure*}[hbtp]
\centering
\includegraphics[trim={0.0cm 0cm 0cm 0cm},clip, width=1\textwidth]{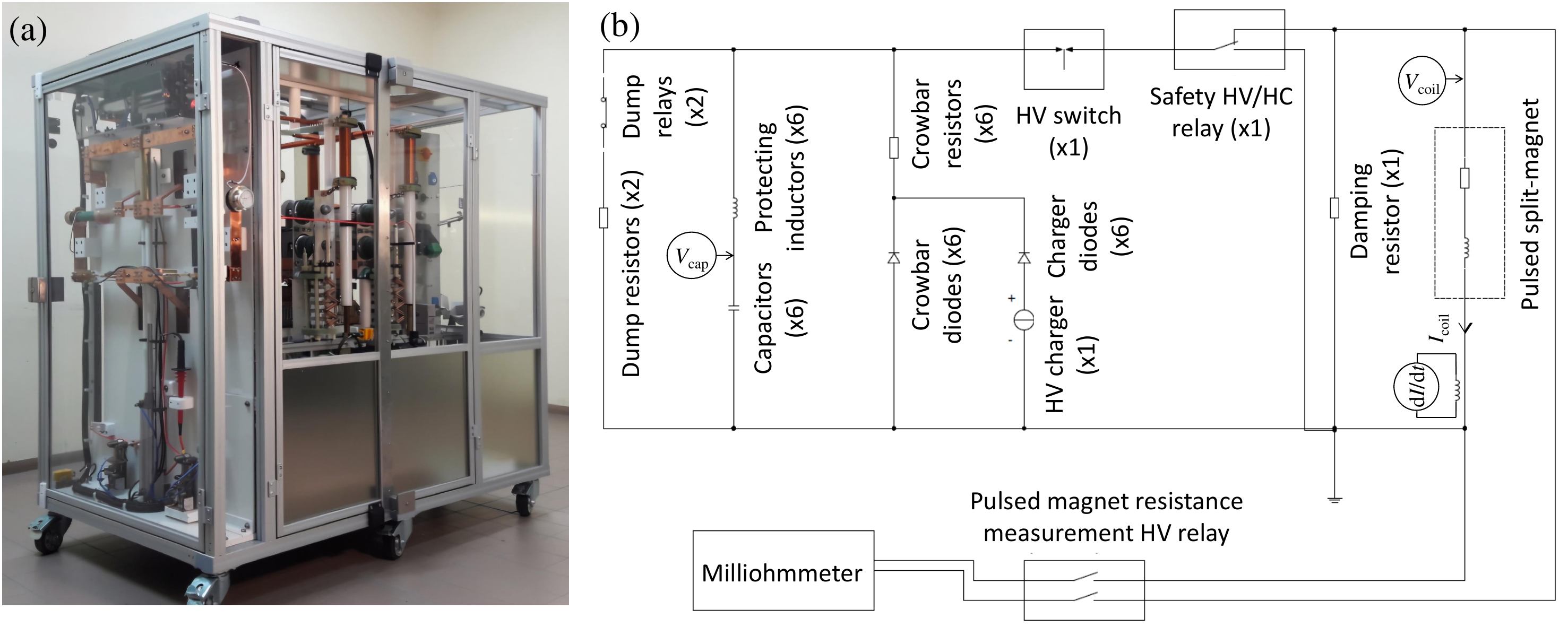}
\caption{\label{pulser} (a) Photo of the capacitor bank (b) Electrical schematic of the generator circuit, highlighting the main components. The numbers in parentheses indicate the number of elements in parallel. High-voltage probes placed on the capacitors and the coil, along with a pickup coil acting as a simplified Rogowski coil, enable monitoring of the system’s proper operation.}
 
 \end{figure*}

 To recollimate the electron beam exiting the selector, the beam is  sent into a pulsed magnetic field \cite{Bolaos2019} that allows to focus the electron beam, and hence to maintain a high-energy-density beam over a long (> 10 cm) distance. 
 The pulsed magnetic field is produced by coupling a high-strength Helmholtz coil \cite{Albertazzi2013} 
 to a pulse-power unit. 
 The latter is a new pulsed power generator, shown in Fig.~\ref{pulser}a, developed by the LNCMI-Toulouse Pulsed-Magnets and Generators group \cite{Moraine2022}. It is a compact and mobile system designed to generate intense magnetic fields in high-power laser environments. Its core consists of a bank of $6~\times~50~\mu F$ capacitors, chargeable up to $24$~kV, storing a maximum energy of 86.4~kJ. With a circuit inductance of 3~$\mu H$ and a resistance of 6~$m\Omega$, it delivers a short-circuit current of up to 220~kA and achieves a minimum discharge rising time of 45~$\mu s$. The connected electromagnets, featuring inductance of a few tens of $\mu H$, provide a typical rise time of $\sim$200~$\mu s$, ensuring a steady-state magnetic field at the nanosecond timescale of laser-plasma interactions. The electrical schematic of the generator is shown in Fig.~\ref{pulser}b.
 
 The system’s energy capacity enables the generation of magnetic fields up to 40~T over several cubic centimeters in room-temperature split coils, with a repetition rate of up to 20 pulses per hour, fitting most high-energy laser facilities. Its modular high-power circuit includes capacitors, a crowbar diode assembly, crowbar resistors, and a limiting inductor. These modules can be selectively connected or disconnected to tailor both energy output and pulse duration for various pulsed split magnets.

 For safety, high-voltage relays automatically discharge the capacitor bank through dump resistors after each pulse or in case of a fault. A PLC-based control system ensures precise operation, while a safety loop equipped with safety relays guarantees seamless integration with the host laboratory’s safety infrastructure. Additionally, a pneumatic high-voltage switch connects the coil to a milliohmeter between pulses to monitor its temperature. The PLC authorizes the next pulse only once the coil has cooled sufficiently.

 Magnet monitoring is performed via oscilloscope acquisition of current variation ($dI/dt$), coil voltage ($V_{coil}$), and capacitor voltage ($V_{cap}$), enabling immediate experiment stop in case of a fault.
 
 The generator is housed in a mobile, aluminum-and-polycarbonate structure ($2015 \times 1220 \times 1940$~$mm^3$, 1.4~t, on 6 wheels), with high-voltage outputs on both sides for flexible integration into different high-power laser experimental room configuration. The split coil is connected via a 70~$mm^2$ coaxial cable and a Jack-type connector, ensuring high-voltage isolation and plug-and-play operation. A visible-grounding, high-voltage/high-current pneumatic switch short-circuits the capacitor bank and disconnects the coil from the generator, guaranteeing operator safety between pulses.

The split-magnet coil that is coupled to the generator has been developed, using the experience LNCMI has had for over a decade in producing split-pair pulsed electromagnets for laboratory astrophysics. These magnets, which employed $\text{CuAg}_\text{0.08\%wt}$ alloys with an ultimate tensile strength (UTS) around 450~MPa, successfully operated at fields up to 30~T. A breakthrough to 40~T was achieved using a $\text{Cu}_\text{60\%}/\text{stainless steel}_\text{40\%}$  macrocomposite conductor, developed in-house at LNCMI \cite{Dupouy1995} to balance high mechanical strength (UTS > 1000~MPa) and high conductivity (60\% IACS).

In conventional high-field pulsed solenoids, each conductor layer is individually reinforced with Zylon fiber (UTS > 5.2~GPa) \cite{vanBocktal1991}, impregnated with Stycast 1266 epoxy, and pre-stressed to more than 700~MPa during winding to achieve a fiber filling factor exceeding 85\% \cite{vanBocktal1991}. This distributed reinforcement technique ensures that hoop stresses in the Zylon layers remain below 3.5~GPa (70\% of UTS) even at peak field, while the copper-based conducting portion of the wire operates in the plastic regime (strain < 1.5\%). Thus, the hoop stress (in a the mid-plane of a cylindrical coil $\sigma_\Phi=j\cdot B(r)\cdot r$) is contained by both the wire and the Zylon layers.

The split-coil geometry (often referred to as Helmholtz coils), \textit{i.e.} two independent coils separated by a 15~mm mid-plane plate addresses key challenges. The system features three optical accesses: first, the axial access (parallel to the magnetic field) has a 13~mm diameter in the 30~T magnet used in the present experiment. Additionally, two radial bores with a 10~mm diameter are integrated into the G11 glass-fiber epoxy composite mid-plane plate. Both can be exchanged by rotating the coil chamber to accommodate the main laser beam, target holders, or diagnostics.

In split coils, the attractive force between the sub-coils becomes so high that the mechanical parts in the mid-plane, where perpendicular accesses are managed, risk collapsing. Due to the short pulse duration, metal mid-plane plates are prohibited and are instead replaced by the G11 insulating composite. Peak compressive stress can exceed 200~MPa at 30~T, a very high value for this material. For high vacuum compatibility, the coils remain in air while the optical accesses are under the laser chamber vacuum. Ensuring the hermeticity of this composite material while subjected to repetitive strong compression is also challenging. Compression tests on various mid-plane plate geometries and optimized layer transitions have been performed, confirming that the mechanical limit can exceed the one reached at the 40~T now routinely achieved.

The coil is designed to be as compact as possible while maximizing mechanical stress. Typical dimensions for 30 or 40~T magnets are roughly: 15~mm inner diameter, 70~mm outer diameter, and 18~mm length per sub-coil, requiring 20–40~kJ to generate their maximum field. Split-coils generating more than 40~T, to be developed in the coming years, will feature smaller optical accesses (inner diameter <10~mm), with slightly larger outer diameter and length, and will utilize the more energy.
Thanks to the new pulsed power supply, the use of liquid nitrogen instead of air for cooling, and the adoption of a $\text{Cu}_\text{40\%}/\text{stainless steel}_\text{60\%}$ macrocomposite conductor (UTS > 1400~MPa at 77~K), magnetic fields approaching 60~T in split coils will soon be available accepting a slightly reduced experimental space of <1~cm in diameter.

\begin{figure}[hbtp]
\centering
\includegraphics[ width=0.45\textwidth]{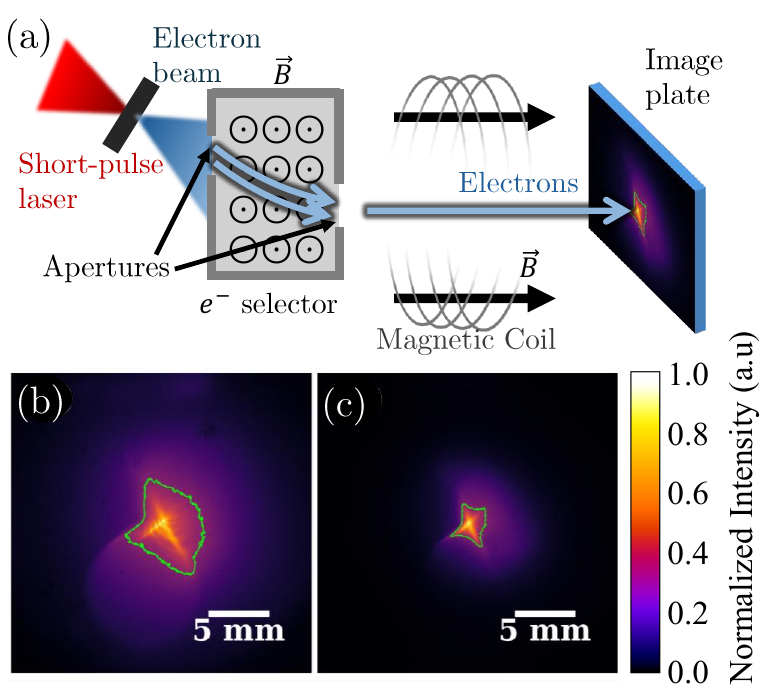}
\caption{(a) Sketch of the coupling of the pulsed magnetic coil to the passive selector in order to collimate the electrons exiting the selector. (b) and (c) Detected electron beam profile after passing through the pulsed magnetic field, of strength 5~T and 10~T, respectively, at the same location as in Fig.\ref{fig:RT_Simulation}b, i.e. 11 cm away from the exit of the selector. }
\label{fig:profiles}%
\end{figure}

 The global setup of the selector, coupled to the split-coil that serves as a beam collimator, is  illustrated in Fig.\ref{fig:profiles}a. The collimation capability of the pulsed magnet in term of beam profile is shown in Fig.\ref{fig:profiles}b-c for two different strengths of the pulsed magnetic field. To quantitatively analyze the beam's spatial profile, we look at 
 the $1/e$ boundary (in green) of the beam. It outlines the high-density core of the electron beam. The effect of the pulsed magnetic field is dramatic, as it allows to reduce the beam spot from 713 mm$^2$ for when uncollimated (see Fig.\ref{fig:RT_Simulation}b) to 30 (6) mm$^2$, for the cases of 5 (10)~T as shown in Fig.\ref{fig:profiles}b (c). The total charge of electrons within the delimitated region was calculated  using the calibration of Ref.~\cite{Bonnet2013}. The average charge pershot within the $1/e$ boundary has been found to be $0.73\pm0.44$~nC, regardless of the use of the magnetic pulser. The error calculated is the standard deviation over 
 multiple shots performed under the same conditions.

\section{Potential applications}

\begin{figure}[hbtp]
\centering
\includegraphics[trim={0cm 0cm 0cm 0cm},clip, width=0.5\textwidth]{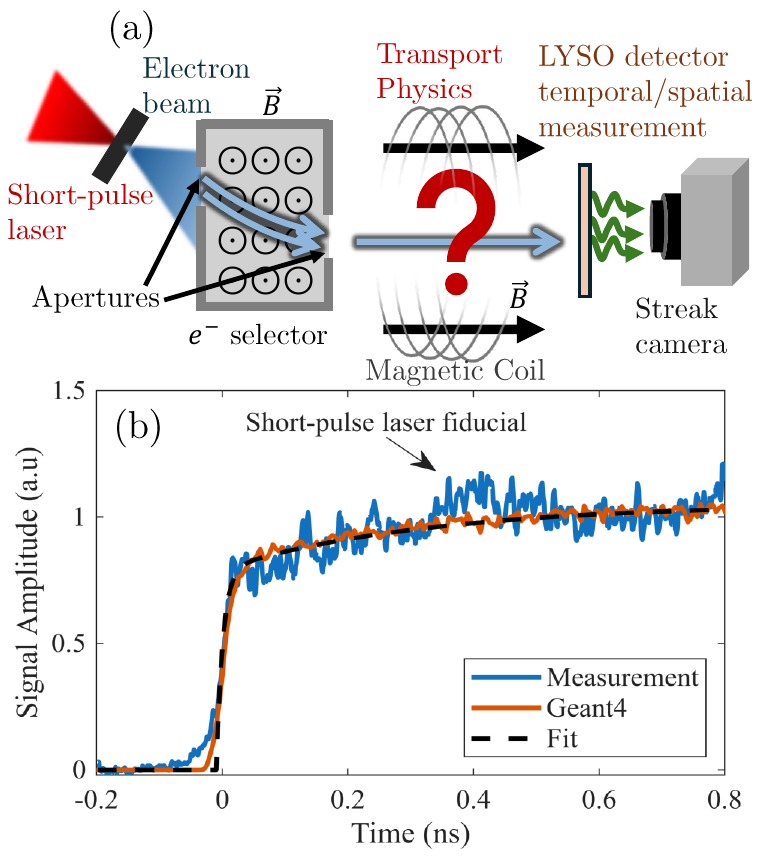}
\caption{ Application of the device detailed in the paper to study electron transport physics, through the  measurement of time-resolved scintillation induced by the electrons after having crossed an expanse of space. (a) cartoon of the setup where an ultrashort burst of energy-selected and collimated electrons 
is sent to interact with a medium, the transport physics we want to investigate. After exiting the medium, the electrons are sent  onto a  LYSO scintillator having  a fast rise time. The emitted light is imaged onto a streak camera. (b) Measured signal in the output of the streak camera, when transporting the electron beam in vacuum under an ambient field of 5T. 
The measured result shows a scintillation double rise time of 10.3~ps and 326~ps, with abundances of 74.9\% and 25.1\%, respectively.}
\label{fig:scintillation}
\end{figure}

Several potential applications may benefit from this electron selector. 
A first is in using the ps-duration electron beam as a probe. This is illustrated in Fig.~\ref{fig:scintillation}a. Such setup would allow to measure the scattering of electrons in plasmas that would be positioned along the trajectory of the electrons. Measuring scattering of high-energy particles in plasma is e.g. a topic of high importance to assess the transport of   cosmic rays (CRs) in space \cite{Kempski2022}. CRs are high-energy particles, that travel through space. CRs strongly interact with the magnetically turbulent interstellar medium (ISM), and this occurs from their birth up to when they straggle within the galaxy. As a consequence, CRs exchange momentum and energy with the ISM, shaping the evolution of astrophysical systems, from stars \cite{Gabici2022} to galaxies as a whole \cite{Ruszkowski2023}. However the exact mechanisms at play still elude us. 

The  setup shown in Fig.~\ref{fig:scintillation}a would effectively allow us to measure  the  transport of electrons in magnetically turbulent laboratory plasma reproducing the astrophysical ones \cite{Cohen2026Turb}, and allow to test models of transport \cite{Kempski2025}. Note that the advantage of using here electrons is that they are relativistic particles, unlike laser-driven laser-driven protons, which are at present limited to hundreds of MeV at best, and that they can  be in the magnetized (small gyro-radius) limit, also unlike protons. To do this, one can record the delay imparted on the electrons by the scattering in the plasma. To measure such delay, we can use an ultrafast scintillator and measure the output scintillation light. Here, we tested using  a LYSO scintillator \cite{Gundacker2016}. This scintillator is well suited to measure electrons in the MeV-range energies, and is the one with 
the  shortest scintillation rise time (10.3~ps, see Fig.~\ref{fig:scintillation}b). The light emitted by the scintillator (peaking at 420~nm) is time-resolved using a visible streak camera (S20 Hamamatsu). The 2D detection plane of the streak camera  yields two pieces of information: (i) the arrival time of the electrons in one dimension, and (ii) the spatial extent occupied by the electron beam in the other dimension. An absolute temporal reference, using a pick-oﬀ of the short-pulse laser used to generate the electrons, imaged onto the streak camera, allows to remove the jitter induced by the electronics of the streak camera.

The measured scintillation signal has been fitted to two rise-time components, see Fig.~\ref{fig:scintillation}b. The rise times have been found to be 10.3~ps and 326~ps, with an abundance of 74.9\% and 25.1\%, respectively, which is  very good agreement with the measurements reported in Ref.~\cite{Gundacker2018}. Fig.~\ref{fig:scintillation}b also shows the scintillation signal simulated using Geant4 \cite{AGOSTINELLI2003250}. We should note here that the scintillator rise time and abundance we used in the simulation are taken from our fit to the measured results. The Geant4 simulation is set to reproduce the full setup: it is initialized with a 2 MeV mono-energetic beam that originates from the  output slit of the passive selector, having an angular spread of 10 degrees (i.e. consistent with the measurement shown in Fig.\ref{fig:RT_Simulation}b) that enables it to cover the full entrance opening of the pulsed coil. The beam then is propagated through the coil system, which operates at a magnetic field of 5 T, as in the measurement, onto the LYSO scintillator. It is useful to note that simulated light output of the scientillator in the simulation is the same, whether we consider 
only the beam and the scintillator, or the full assembly. In other words, the collimating pulsed coil served to collimate the beam, but does not influence its temporal dynamics. 
The simulation result is in   excellent agreement with the measurement, confirming that the passive selector outputs an ultra-short electron burst with a $\sim$ ps temporal spread. With this setup, we show that we are capable of measuring tens of ps delays that would be induced by the scattering of the electrons is a plasma that would be positioned within the coil.

Another application of the device developed here would lie in the domain of medical applications. Indeed, applications such as FLASH-RT delivery \cite{Jo2023,Moreau2025} require the delivery of ultra-short, high-charge electron bunches at high doses in a fraction of a second, triggering the FLASH effect to spare healthy tissue while killing cancer cells. In this frame, the device detailed here  could also benefit such application. Indeed, the selector offers higher beam pointing stability, spectral stability, and simple tuneable control compared to wakefield-based accelerators.

\section{Summary}
A compact passive energy-selector for MeV-range electrons produced by irradiating solid targets with ultra-intense short-pulse lasers has been realized. The device, the tunability of which is based on varying the strength of the magnetic field the electrons travel through (by adjusting the  gap between the magnets in the assembly), has been used to select electron energies in the range of 0.5-5 MeV with excellent agreement with the analytical calculation. The selected beam is then sent through a pulsed magnetic-field system to reduce its transverse divergence. The full system offers perspectives for several applications, from electron probing of plasma, e.g. in the frame of laboratory astrophysics, to 
flash radiation therapy.

\begin{acknowledgments}
We wish to acknowledge the expert support of the technical teams of LULI2000 in the realization of the experiment.
I.S. acknowledges the support from the project ELI-RO/RDI/DEZ/DFG/2025\_032 funded by the Romanian Ministry of Education and Research.
J.F. acknowledges the support of ISF Grant No. 693/26.
\end{acknowledgments}

\section*{Data Availability Statement}

The data that support the findings of this study are available from the corresponding author upon reasonable request.


\bibliography{MainBib}

\end{document}